\documentclass[aps,twocolumn,prl,superscriptaddress]{revtex4-1}

\usepackage{color}
\usepackage{verbatim}
\usepackage{hyperref}
\usepackage{graphicx}
\usepackage{dcolumn}
\usepackage{bm}

\begin{document}
\title{Interacting Electron Wave Packet Dynamics in a Two-Dimensional Nanochannel}

\author{Christoph M. Puetter}
\email{cpuetter@comas.frsc.tsukuba.ac.jp}
\affiliation{Graduate School of Pure and Applied Sciences, University of Tsukuba, 1-1-1
  Tennodai, Tsukuba, 305-8577, Japan}
\affiliation{CREST, Japan Science and Technology Agency, 7 Gobancho, Chiyoda, Tokyo 102-0075, Japan}
\author{Satoru Konabe}
\affiliation{Graduate School of Pure and Applied Sciences, University of Tsukuba, 1-1-1
  Tennodai, Tsukuba, 305-8577, Japan}
\affiliation{CREST, Japan Science and Technology Agency, 7 Gobancho, Chiyoda, Tokyo 102-0075, Japan}
\author{Yasuhiro Hatsugai}
\affiliation{Graduate School of Pure and Applied Sciences, University of Tsukuba, 1-1-1
  Tennodai, Tsukuba, 305-8577, Japan}
\affiliation{CREST, Japan Science and Technology Agency, 7 Gobancho, Chiyoda, Tokyo 102-0075, Japan}
\affiliation{Center for Interdisciplinary Research, Tohoku
  University, Sendai 980-8578, Japan},
\author{Kenji Ohmori}
\affiliation{Graduate School of Pure and Applied Sciences, University of Tsukuba, 1-1-1
Tennodai, Tsukuba, 305-8577, Japan}
\affiliation{CREST, Japan Science and Technology Agency, 7 Gobancho, Chiyoda, Tokyo 102-0075, Japan}
\author{Kenji Shiraishi}
\affiliation{Graduate School of Pure and Applied Sciences, University of Tsukuba, 1-1-1
Tennodai, Tsukuba, 305-8577, Japan}
  \affiliation{CREST, Japan Science and Technology Agency, 7 Gobancho,
    Chiyoda, Tokyo 102-0075, Japan}
\affiliation{Center for Computational Sciences, University of Tsukuba, 1-1-1 Tennodai,
Tsukuba, 305-8577, Japan}

\begin{abstract}
  Classical and quantum dynamics are important limits for the
  understanding of the transport characteristics of interacting
  electrons in nanodevices.
  Here, we apply an intermediate semiclassical approach to investigate the dynamics
  of two interacting electrons in a planar nanochannel as a function of
  Coulomb repulsion and electric field.
  We find that charge is mostly redistributed to the channel edges 
  and that an electric field enhances the particle-like character of electrons. 
  These results may have significant implications for the
  design and study of future nanodevices.
\end{abstract}

\maketitle

Ideal ultrasmall logic nanodevices feature high switching speeds, low power
consumption, excellent on-off current ratios and high scalability
and integrability.
Electron transport in ultrasmall nanodevices approaching 
channel lengths of approximately $10 \, \text{nm}$, 
\cite{ITRS2011}
however, turns out to be quasi-ballistic and intricate 
\cite{Sano2011JJAP,Sano2011JComptElectron}
due to Brownian motion (thermal noise
caused by scattering and diffusion) and the 
discreteness of the electric charge (leading to shot noise 
enhanced by unscreened trapped charges).
These effects give rise to significant current  
fluctuations at high clock speeds and low voltages,
\cite{Kamioka2012IEDM,Feng2012IEDM} which are
detrimental to efficient device operation.
This indicates that high-speed electrons in nanoscale regions cannot be
described by conventional statistical frameworks such as Maxwell-Boltzmann approaches. 
Highly doped drain and source regions can further impact channel
electrons, e.g., due to the build-up of mirror charges, 
suggesting that Coulomb interaction is a paramount ingredient 
in describing transport, dissipation, and equilibration in nanostructures. 
\cite{Sano2011JJAP,Sano2011JComptElectron,Lee2009APL} 
A comprehensive understanding of electron dynamics on small time and
length scales therefore is desirable to improve the device performance.

\begin{figure}[t!]
  \centering
  \includegraphics*[width=1.0\linewidth, clip]{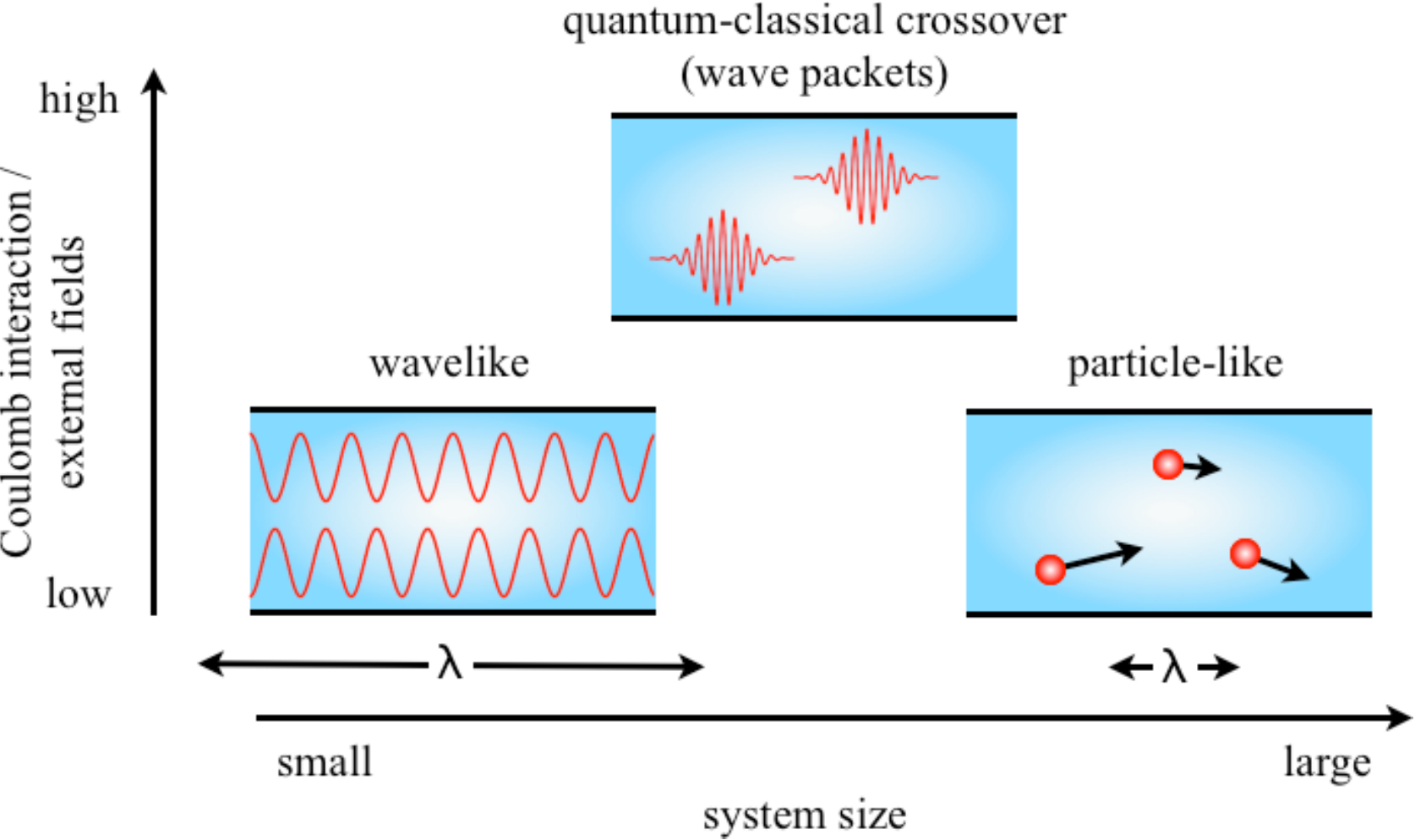}
  \caption{Schematic illustration of 
    classical (particle-based), quantum (wave-based) and intermediate
    electronic transport regimes applicable to a nanochannel.
    For intermediate system sizes a semiclassical approach based on 
    a wave packet picture may be used, which is further strengthened
    by an external field and/or Coulomb interaction.
    ($\lambda$ indicates the electron mean free path.)
    \label{fig:SemiclassicalXoverRegime.001cropped}}
\end{figure}

Various approaches to electron transport in nanodevices 
have been taken so far, ranging from classical 
Monte Carlo and molecular dynamics methods to 
quantum nonequilibrium Green function calculations. 
\cite{Kamioka2012IEDM,Feng2012IEDM,Sano2011JJAP,DattaBook,HaugJauhoBook}
Here, we address the charge transport from an intermediate, 
semiclassical perspective,
\cite{Xiao2010RMP,Shindou2005NuclPhysB}
by solving the Schr\"{o}dinger equation numerically
for a pair of interacting electron wave packets that
propagate in a planar nanochannel.
This approach interpolates between the classical, particle-like and 
the quantum, wavelike nature of electrons
(Fig. \ref{fig:SemiclassicalXoverRegime.001cropped})  and approximately 
preserves the discreteness of the transported electron charge over 
appropriate (not too short) time scales.  
Wave packet approaches have been considered previously,
but relied on strictly one-dimensional structures and/or on 
noninteracting electrons, or were based on 
approximate schemes such as the time-dependent Hartree-Fock theory.
\cite{Fu2005JApplPhys,Claro2003PRB,Dias2007PRB,Pruneda2009PRB,Kramer2010PhysScr,deSousa2012JAP,Shiokawa2013JJAP}

Below, we study in detail the effect of Coulomb repulsion
and an external electric field on wave packet 
propagation in a two-dimensional nanochannel.
A main finding is that Coulomb repulsion redistributes the
charge density to the channel walls, which
makes electron transport more sensitive to 
perturbations at the interface.
The presence of a uniform electric field, mimicking a channel
potential, furthermore has a stabilizing effect 
on the wave packets, reducing the spreading 
along the channel direction and, hence, backing 
the more localized, classical particle-like picture often used in full-scale
device simulations.

\begin{figure*}[t!]
  \centering
  \includegraphics*[width=1.0\linewidth, clip]{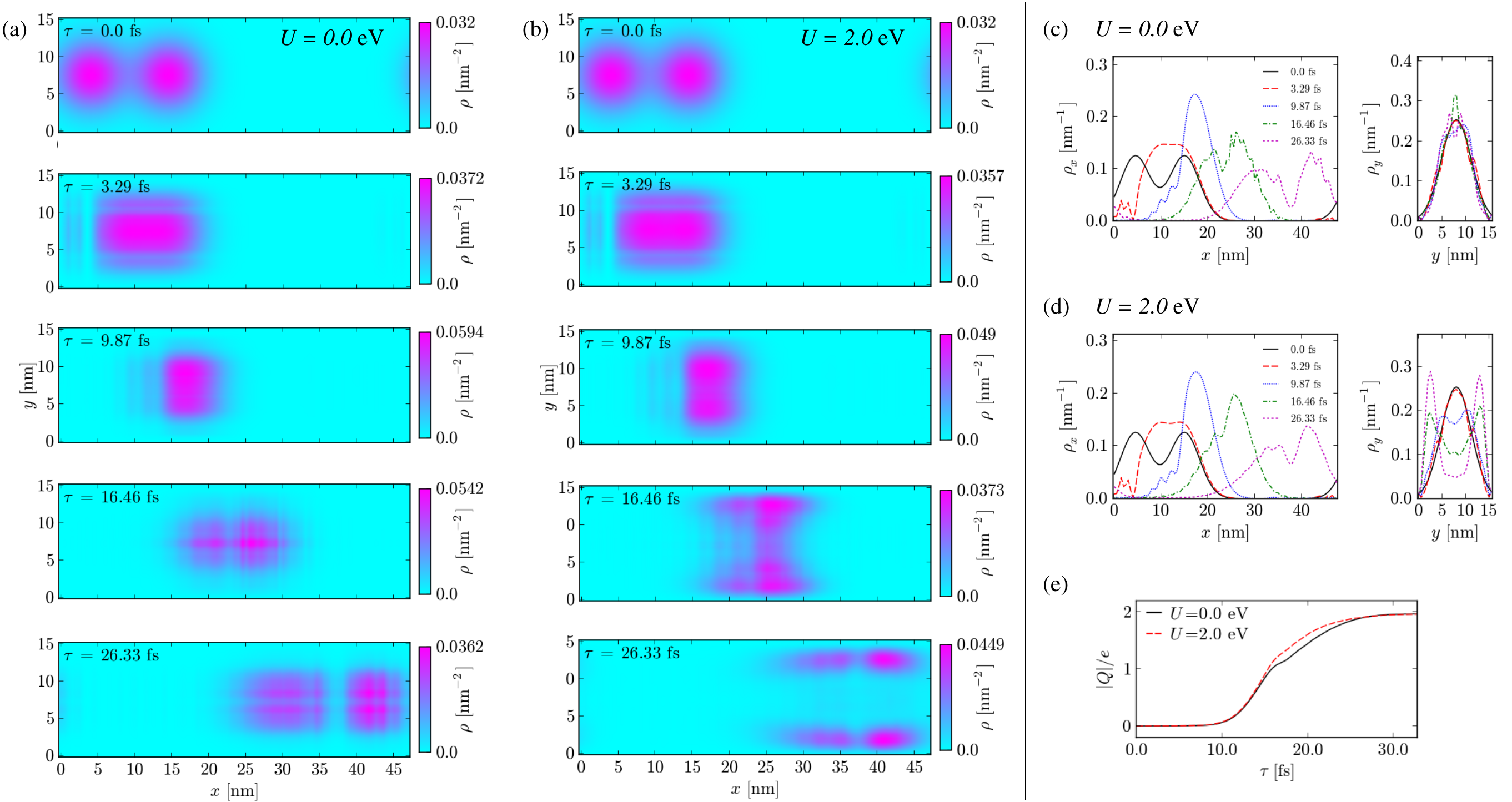}
  \caption{Comparison of the time evolution of two noninteracting and 
    two interacting electron wave packets in a two-dimensional
    nanochannel in the presence of an electric field applied in x-direction.
    (a), (b) Snapshots of the electron density $\rho({\bf
      r}; \tau)$ for $U=0$ and $2.0$ eV, respectively.
    (c), (d) Corresponding partially integrated one-dimensional
    electron densities
    $\rho_{x}(x; \tau)$ and $\rho_{y}(y; \tau)$ for $U=0$ 
    and $2.0$ eV, respectively.
    (e) Total charge transported across the half-way line at $x_{0}=23.5$ nm as a
    function of time $\tau$.
    The initial sizes of the Gaussian wave packets are 
    $\sigma_{r} = 9 a = 4.5$ nm, and the electric field 
    amounts to $E=0.4$ MV/cm; only the incoming, initially left-hand side wave packet
    possesses initially a finite momentum ${\bf k}=(0.3 \pi/a, 0)$.
    \label{fig:outTwo000131ca000133c.1D2DSnapshotswTC}}
\end{figure*}

In order to study the propagation of two interacting electrons in a
nanochannel, we solve the time-dependent Schr\"{o}dinger equation 
$i \hbar \partial_{\tau} | \psi(\tau) \rangle = H(\tau) | \psi(\tau)
\rangle$ 
numerically using a split-operator
(Suzuki-Trotter) approach.
\cite{Suzuki1990PhysLettA,Suzuki1992JPSJ,Hatsugai2001IntJModPhysB}
The total Hamiltonian $H(\tau) = H_{\text{kin}}(\tau) +
H_{\text{int}}$ includes a kinetic and a Coulomb interaction term, i.e.,
\begin{eqnarray}
  H_{\text{kin}}(\tau) &=& -t \sum_{{\bf r}} (\text{e}^{- \frac{i e}{\hbar}
    E a \tau} c^{\dagger}_{{\bf r}} c_{{\bf
      r}+{\bf a}_{x}} + c^{\dagger}_{{\bf r}} c_{{\bf
      r}+{\bf a}_{y}} + \text{h.c.}) \\
  H_{\text{int}} &=& \frac{1}{2} \sum_{{\bf r}, {\bf r}'} \frac{U}{|{\bf r} -
    {\bf r}'|/a} c^{\dagger}_{{\bf r}} c_{{\bf r}} c^{\dagger}_{{\bf r}'}
  c_{{\bf r}'},
\end{eqnarray}
respectively, where $c^{\dagger}_{{\bf r}}$ ($c_{{\bf r}}$) creates
(annihilates) a spinless electron at site ${\bf r}$.
Here, the kinetic Hamiltonian $H_{\text{kin}}(\tau)$
describes nearest-neighbor hopping 
processes with hopping amplitude $t$ and is time-dependent 
through a Peierls phase factor to account for a uniform 
electric field ${\bf E}=(E, 0)$ in x-direction.
By coupling the electrons to an electric field in this way, the channel potential 
along the x-direction can be modelled without
spurious discontinuities along x associated with the finite size 
of a periodic channel (at the expense of the conservation of
total energy), whereas closed boundary conditions are
imposed along y.
The lattice spacing and two lattice vectors of the underlying
square lattice are denoted by $a$, ${\bf a}_{x}$ and ${\bf a}_{y}$,
respectively. 
The term $H_{\text{int}}$ represents the long-range Coulomb
interaction with interaction strength $U$.

\begin{figure*}[t!]
  \centering
  \includegraphics*[width=1.0\linewidth, clip]{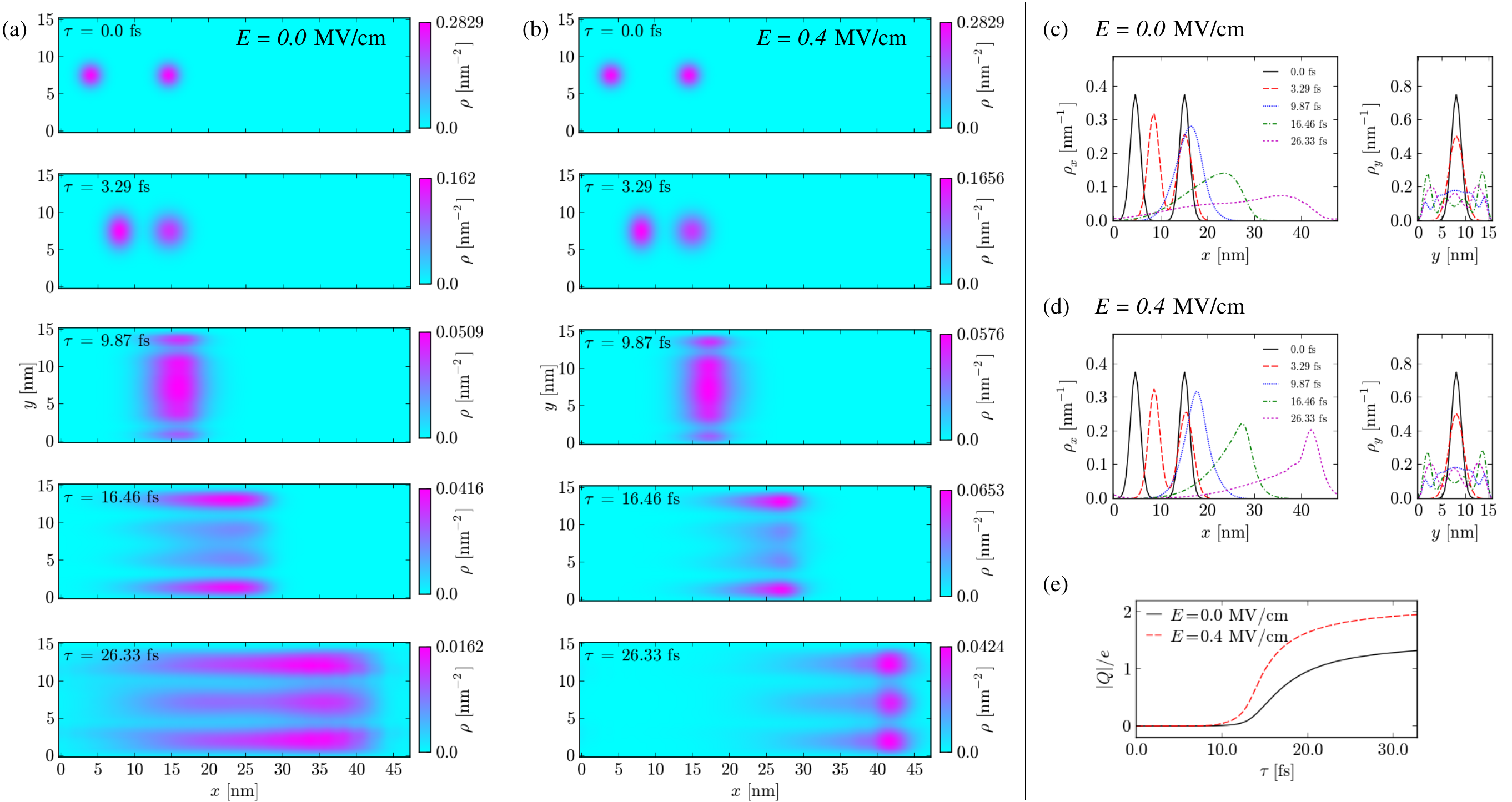}
   \caption{Effect of an electric field applied in x-direction
    for a fixed Coulomb repulsion magnitude.
    (a), (b) Snapshots of the electron density $\rho({\bf
      r}; \tau)$ for $E=0$ and $0.4$ MV/cm, respectively.
    (c), (d) Corresponding partially integrated one-dimensional
    electron densities
    $\rho_{x}(x; \tau)$ and $\rho_{y}(y; \tau)$ for $E=0$ and $0.4$ MV/cm, respectively.
    (e) Total charge transported across the half-way line at $x_{0}=23.5$ nm as a
    function of time $\tau$.
    The initial wave packet sizes are given by
    $\sigma_{r} = 3 a = 1.5$ nm, while $U=2.0$ eV;
    only the initially left-hand side wave packet
    is initially given a finite momentum ${\bf k}=(0.3 \pi/a, 0)$.
    \label{fig:outTwo000113a000133a.1D2DSnapshotswTC}}
\end{figure*}

We are interested in the evolution of
the two-particle state $| \psi(\tau) \rangle =(1/\sqrt{2}) \sum_{{\bf r}, {\bf r}'}
\Psi({\bf r}, {\bf r}'; \tau) c^{\dagger}_{{\bf r}} c^{\dagger}_{{\bf
    r}'} |0\rangle$, 
where $\Psi({\bf r}, {\bf r}'; \tau)$ stands for the antisymmetric
two-particle real space wave function at time $\tau$,
normalized to $\sum_{{\bf r}, {\bf r}'} |\Psi({\bf r}, {\bf r}';
\tau)|^{2} = 1$. 
With $L_{x} a$ and $L_{y} a$ denoting the extend of the lattice in x
  and y direction,  the periodic and closed boundary conditions take the form $\Psi({\bf
    r}, {\bf r}'; \tau) = \Psi({\bf r} \pm L_{x} {\bf a}_{x}, {\bf r}';
  \tau) = \Psi({\bf r}, {\bf r}' \pm L_{x} {\bf a}_{x}; \tau) =
  \Psi({\bf r} \pm L_{x} {\bf a}_{x}, {\bf r}' \pm L_{x} {\bf
    a}_{x}; \tau)$
and 
$\Psi({\bf r}, {\bf r}'; \tau) = 0$ if $r_{y}, r'_{y} \geq
L_{y} a$ or $< 0$ [${\bf r} = (r_{x}, r_{y})$, ${\bf r}' = (r'_{x}, r'_{y})$], respectively.
As an initial condition, we model the probability distribution of the 
electrons by two (moving) Gaussian wave packets centered at ${\bf R}$ and
${\bf R}'$ with momenta ${\bf k}$ and ${\bf k}'$ and width $\sigma_{r}$,
\begin{eqnarray}
  \Psi({\bf r}, {\bf r}'; 0) &=& \frac{1}{\mathcal{N}} [\Phi_{{\bf k}}({\bf
    r}-{\bf R})  \Phi_{{\bf k}'}({\bf r}'-{\bf R}') \nonumber \\
  && \hspace{1.5cm} - \Phi_{{\bf k}'}({\bf
    r}-{\bf R}')  \Phi_{{\bf k}}({\bf r}'-{\bf R})],
\end{eqnarray}
where $\Phi_{{\bf k}}({\bf r}) \propto \text{exp}[-|{\bf r}|^{2}/(2
\sigma^{2}_{r}) + i {\bf k} \cdot {\bf r}]$
and $\mathcal{N}$ ensures proper normalization.
Below, we consider only scenarios where the
right-hand side wave packet is initially at rest (${\bf k}'=0$) 
while the left-hand side wave packet initially possesses 
a finite incoming crystal momentum centered at ${\bf k}=(0.3 \pi/a, 0)$]
(see the snapshot sequences in
Figs. \ref{fig:outTwo000131ca000133c.1D2DSnapshotswTC} and
\ref{fig:outTwo000113a000133a.1D2DSnapshotswTC}).

For a convenient but adequate set of parameters relevant to
semiconducting nanodevices, we choose $t=1$ eV, which corresponds to an
effective band mass of $m^{*}= 0.763 \, m_{e}$ in units of free
electron mass $m_{e}$, and $a = 0.5$ nm for the lattice spacing.  
Since the channel lengths of nanodevices will reach only a few tens of
nm in the near future, we further base the wave packet simulation
on a lattice of size $48 \times 16$ nm$^{2}$ 
(corresponding to $L_{x} \times L_{y} = 96 \times 32$ lattice sites), representing a
rather long channel to reduce finite size effects.

The typical time evolution of a wave packet pair in the presence of a
moderate finite electric field $E=0.4$ MV/cm
and in the absence ($U=0$) or presence ($U = 2.0$ eV) 
of Coulomb interaction is summarized in
Fig. \ref{fig:outTwo000131ca000133c.1D2DSnapshotswTC}, where 
$\sigma_{r} = 4.5$ nm.
The snapshots in
Figs. \ref{fig:outTwo000131ca000133c.1D2DSnapshotswTC}(a) 
and \ref{fig:outTwo000131ca000133c.1D2DSnapshotswTC}(c) for $U=0$, which display the electron density 
$\rho({\bf r}; \tau) = 2 \sum_{{\bf r}'} |\Psi({\bf r}, {\bf r}';
\tau)|^{2}$ and 
the corresponding one-dimensional electron densities
 $\rho_{x}(x; \tau) = \sum_{y} \rho({\bf r}; \tau)$ 
and $\rho_{y}(y; \tau) = \sum_{x} \rho({\bf r}; \tau)$
[${\bf r}=(x, y)$] along the x- and y-directions, respectively,
clearly imply that the evolution of the wave packets is strongly determined
by the lateral confinement in the y-direction and to a lesser extent by the Pauli exclusion
principle (since we are considering spinless electrons only).
Due to the confinement in the y-direction, the spreading of the wave
packets and the subsequent reflection at the channel walls lead to
interference along y, e.g., 
causing a pronounced temporary maximum 
on the channel centre line at $\tau = 16.46$ fs.
The partial reflection of the incoming wave packet off the initially static
wave packet due to Pauli exclusion also causes an, albeit much weaker,
interference pattern along the x-direction.
Most of the incoming wave packet, however, 
passes by the initially static wave packet such that both 
wave packets remain essentially distinct and intact, propagating with
different and increasing velocities.
Under the present electric field, 
the peak-to-peak amplitude of the corresponding 
Bloch oscillation amounts to $2 s_{0} = 4 t/(e E) =
100$ nm, which substantially exceeds the channel length. \cite{Hartmann2004NJP}

The effect of finite Coulomb repulsion ($U=2$ eV) is clearly visible in the
snapshot sequences of 
Figs. \ref{fig:outTwo000131ca000133c.1D2DSnapshotswTC}(b) 
and \ref{fig:outTwo000131ca000133c.1D2DSnapshotswTC}(d).
Due to the Coulomb interaction, the incoming wave packet pushes the
initially resting wave packet forward and enhances the transverse expansion
along the y-direction.
The latter results in a more pronounced interference pattern along y,
with the overall probability weight of the wave packets pushed further
toward the channel edges. 
The Coulomb interaction also slightly enhances the charge transport along 
the longitudinal channel direction, 
as can be seen from the total charge transported
across the line at $x_{0}=23.5$ nm
[see Fig. \ref{fig:outTwo000131ca000133c.1D2DSnapshotswTC}(e)].  
The total transported charge $Q(\tau)$  can be obtained by integrating
the probability current density in the x-direction over time and the
channel width. 
The x-component of the current density operator is given by
\begin{eqnarray}
  j^{x}({\bf r}; \tau) = -\frac{i t a}{\hbar} \big[ \text{e}^{-i
    \frac{i e}{\hbar} E a \tau} c^{\dagger}_{{\bf r}-{\bf a}_{x}} c_{\bf r} - \text{h.c.}\big],
\end{eqnarray}
which can be derived from the appropriate continuity equation, \cite{Costa2012PRB}
yielding
\begin{eqnarray}
  Q({\tau}) = (-e) \sum_{y } \int_{0}^{\tau} \langle \psi(\tau') |
  j^{x}(x_{0}, y; \tau') | \psi(\tau') \rangle \text{d}\tau'.
\end{eqnarray}
Furthermore, we note that the time dependence of the
total transported charge seems largely independent of the initial
size, shape (e.g., quasi-one-dimensional Gaussians versus circular Gaussians) 
or positions (centered or slightly off-centered with respect to each
other) of the wave packets.
Finally, the interaction strength $U = 2 t$ used here
seems large, but was choosen to highlight the qualitative impact of
Coulomb repulsion on the wave packet evolution, which we found
is qualitatively similar for $U=t, 2t, 3t$ ($=1, 2, 3$ eV).

The effect of a moderate electric field in the presence of finite Coulomb
repulsion ($U = 2.0$ eV) is displayed in 
Fig. \ref{fig:outTwo000113a000133a.1D2DSnapshotswTC},
which compares the wave packet evolution for $E = 0$ and $0.4$ MV/cm.
Here, we have chosen a pair of initially smaller wave packets with $\sigma_{r}=1.5$ nm,
which spread more quickly with time and generate a more heavily modulated
density distribution in the transverse channel direction 
[cf. panels (c) and (d) of Figs. \ref{fig:outTwo000131ca000133c.1D2DSnapshotswTC}
and \ref{fig:outTwo000113a000133a.1D2DSnapshotswTC}].
The wave function nevertheless maintains a larger weight 
near the channel edges than in the centre, similar to the findings above.

More strikingly, however, the overall expansion of the electron density 
in the x-direction is drastically reduced in the presence of a finite
electric field [cf. the snapshots in
Figs. \ref{fig:outTwo000113a000133a.1D2DSnapshotswTC}(a)-\ref{fig:outTwo000113a000133a.1D2DSnapshotswTC}(d) at time $\tau = 26.33$ fs],
where a density maximum in the x-direction clearly develops within the simulation period.
A similar electric field effect can also be observed in the case of larger initial
wave packets.
The electric field, expectedly, also leads to a significant increase in
the transported charge
[Fig. \ref{fig:outTwo000113a000133a.1D2DSnapshotswTC}(e)].
In the absence of an external field, the transported charge starts to saturate 
near unit charge $|e|$ at about $\tau=20$ fs, indicating that, despite 
the (weakly) current enhancing influence 
of the Coulomb interaction, largely only one electron wave packet
reaches the right half of the channel during the
simulation time.

By modeling the initial electron probability distribution via two interacting wave packets,
the present study takes a semiclassical approach to charge
transport in a planar nanochannel. \cite{Xiao2010RMP,Shindou2005NuclPhysB}
Our simulations show that the uniform spreading of the wave packets
is limited along the transverse channel direction 
due to both interference effects and Coulomb repulsion.
As a result, the electron density is strongly modulated along 
the y-direction, exhibiting a stationary statelike distribution.
However, it propagates along the
channel x-direction with a relatively large probability weight near (but
not too close to) the channel edges.
This density pattern can also be interpreted as the pair of electrons occupying
only the low-lying transverse eigenmodes whose probability distribution is
modified and pushed closer to the channel edges due to the Coulomb
interaction. 
The details and extent of such edge-dominated transport can have significant implications
for the performance of nanodevices, where interface roughness and
charge trapping at interfaces are known to lead to a substantial degradation of
the electron mobility. \cite{Kamioka2012IEDM,Feng2012IEDM}

Furthermore, we find that for typical nanochannel length and time scales,
an electric field has a stabilizing effect on electron
wave packets, inhibiting spreading along the channel direction.
This behavior reinforces the more particle-like
picture of electrons in a nanochannel, lending support
to particle-based classical or semiclassical 
studies such as classical Monte Carlo and/or
molecular dynamics methods to solve 
the classical Boltzmann transport equation in 
the presence of interactions and impurities. 
The stabilizing effect of an electric field has also been observed 
for strictly one-dimensional interacting wave packets
based on a time-dependent Hartree-Fock approach,
which additionally revealed a bunching phenomenon of wave packets within 
the one-dimensional channel. \cite{Shiokawa2013JJAP}

In conclusion, we have considered the effect of  
Coulomb repulsion and external field on the electron dynamics in a nanochannel,
by exactly solving the Schroedinger equation 
for a pair of interacting and propagating electrons.
This is in contrast to previous wave packet studies that
mostly focused on single electrons and/or one-dimensional nanostructures.
\cite{Claro2003PRB,Fu2005JApplPhys,Dias2007PRB,Pruneda2009PRB,Kramer2010PhysScr}
The present wave packet simulation 
may be extended in future studies to include, e.g., the electronic spin degree of
freedom, small three-dimensional nanostructures, more
than two interacting electrons and, though more challenging, 
the coupling to high-density source and drain contacts.

\acknowledgements

The authors thank Y. Tokura for useful discussions.

\end{document}